\documentclass[useAMS,usenatbib]{mn2e}

\usepackage{graphicx}
\usepackage{float}
\usepackage{placeins} 
\usepackage{paralist}

\newcommand{\HI}{H\,{\sc i}}

\newcommand{\Msun}{$M_{\odot}$}
\newcommand{\kms}{km\,s$^{-1}$}
\newcommand{\vLG}{$v_{\rm LG}$}

\title{New \HI\ scaling relations to probe the \HI\ content of galaxies via global \HI-deficiency maps}
\author[H. D\'{e}nes, V. A. Kilborn, B. S. Koribalski]
  {H.~D\'{e}nes$^{1, 2}$\thanks{E-mail: hdenes@astro.swin.edu.au}, V. A. Kilborn$^1$, B. S. Koribalski$^2$
   \\
  $^1$Centre for Astrophysics \& Supercomputing, Swinburne University of Technology\\
  $^2$Australia Telescope National Facility, CSIRO Astronomy and Space Science, P.O. Box 76, Epping, NSW 1710, Australia\\
  }
\date{Released 2002 Xxxxx XX}

\pagerange{\pageref{firstpage}--\pageref{lastpage}} \pubyear{2002}

\def\LaTeX{L\kern-.36em\raise.3ex\hbox{a}\kern-.15em
    T\kern-.1667em\lower.7ex\hbox{E}\kern-.125emX}

\begin{document}

\label{firstpage}

\maketitle

\begin{abstract}

We present new multi-wavelength scaling relations between the neutral hydrogen content (\HI) and the stellar properties of nearby galaxies selected from the \HI\ Parkes All-Sky Survey (HIPASS). We use these new scaling relations to investigate the environmental dependency of the \HI\ content of galaxies. We find that galaxies in high density environments tend to have on average less \HI\ than galaxies with the same stellar mass in the low density environment. Our new \HI\ scaling relations allow us to identify individual galaxies, as well as group/cluster environments, that have an `anomalous' \HI\ content. We map the global distribution of \HI-deficient and \HI-excess galaxies on the sky and compare it to the large scale structure of galaxies. We find galaxy clusters to be \HI-deficient, and we identify that the regions surrounding clusters tend to be \HI-excess. Finally, we demonstrate the potential of using \HI\ scaling relations to predict future \HI\ surveys based on an optical redshift survey. We apply our scaling relations to 16709 galaxies in the 6dF Galaxy Survey (6dFGS) that lie in the HIPASS volume and compare our predictions to the measurements. We find that scaling relations are good method to estimate the outcome of \HI\ surveys. 

\end{abstract}


\section{Introduction}

One of the main challenges of extragalactic astrophysics is to understand the role of cold gas in the formation and evolution of galaxies. Neutral hydrogen gas (\HI) is one of the most important building blocks of galaxies, since it is the main reservoir for future star formation along with molecular hydrogen. Without a significant reservoir of cold gas, star formation terminates and the galaxy becomes red and passive (e.g. \citealt{Larson1980}). 

We know that the environment plays an important role in the gas content of a galaxy and in galaxy evolution. For example the fraction of early-type galaxies increases and the fraction of late-types decreases with increasing galaxy density. This is called the `morphology density relation' \citep{Dressler1980, Fasano2000, Goto2003}. Another important observation is that the fraction of late-type galaxies in galaxy clusters increases with increasing redshift, which is called the Butcher-Oemler effect \citep{Butcher1987}. These effects suggest that environmental processes play an important role in transforming late-type star forming galaxies into passive red early type galaxies over time. 

In terms of \HI\ content, spiral galaxies in high density environments tend to have, on average, less \HI\ than galaxies of the same type and size in the field (e.g. \citealt{Davis1973, Giovanelli1985, Solanes2001}). This implies that late type galaxies in high density regions are getting stripped of, or use up, their gas and are transforming to gas poor early type galaxies (e.g. \citealt{Gunn1972, Fasano2000, Bekki2002, Bekki2011}). Possible gas stripping mechanisms, such as mergers, tidal interactions \citep{Mihos2005}, ram pressure stripping \citep{Gunn1972}, turbulent or viscous stripping \citep{Nulsen1982}, thermal evaporation \citep{Cowie1977}, starvation \citep{Larson1980} and harassment \citep{Moore1996}, are well studied in high density environments, especially in the Virgo cluster (e.g. \citealt{Kenney2004, Vollmer2001, Chung2009}). There is more and more evidence that environment starts to influence galaxy evolution at densities typical of poor groups \citep{Chamaraux2004, Kilborn2005, Sengupta2006, Kilborn2009, Westmeier2011}, but gas stripping mechanisms are not nearly as well studied in low density environments than in galaxy clusters. 

HI-optical scaling relations are a useful tool to investigate how different density environments influence the gas content of galaxies. We can characterise the \HI\ content of galaxies with scaling relations between the \HI\ content and other intrinsic properties of galaxies. Then we can approach environmental effects from two points of view. Firstly we can investigate global, statistical trends between the \HI\ content and the environment density or, secondly we can look at individual galaxies that differ from the average. We can identify such galaxies with anomalous \HI\ content by using scaling relations to calculate their expected \HI\ mass and comparing it to their measured \HI\ mass. Anomalous galaxies may have either less \HI\ than expected (\HI-deficient) or they have more \HI\ than expected (\HI-excess). To quantify the relative gas content of galaxies \cite{Haynes1983} introduced the ``deficiency factor''. The \HI\ deficiency factor (\textsc{DEF}) is expressed as a logarithmic quantity, positive for \HI-deficient galaxies and negative for galaxies with \HI-excess. 
\begin{equation}
 \textsc{Def}_{HI} = \rmn{log[M}_{HI exp}] - \rmn{log[M}_{HI obs}], 
\end{equation}
 where M$_{HI exp}$ is the expected \HI\ mass, usually calculated from \HI\ scaling relations, and M$_{HI obs}$ is the calculated \HI\ mass from the measurements.
 
Previous works investigating HI-optical scaling relations for late type galaxies found that the optical diameter and luminosity correlate well with the \HI\ mass (e.g. \citealt{Haynes1984}; \citealt{Chamaraux1986}; \citealt{Solanes1996}). Initial scaling relations were determined from relatively small, optically selected samples, that have a natural bias against blue, low surface brightness objects, which tend to have significant amounts of \HI. More recent studies favour the use of \HI\ selected samples combined with optical properties from the Sloan Digital Sky Survey (SDSS; \citealt{York2000}). For example \citet{Toribio2011_2} investigated a sample of isolated galaxies from the Arecibo Legacy Fast ALFA (ALFALFA) blind 21 cm line survey \citep{ALFALFA} and found that the best indicator for \HI\ mass is the SDSS \textit{r}-band diameter, followed by the total luminosity and the maximum rotational speed of the galaxy. Another approach to estimate a galaxy's \HI\ content is to use recent star formation indicators such as UV to optical colours. \citet{Catinella2010} showed that the linear combination of \textit{NUV-r} colour and stellar surface density is a good predictor of the gas content of massive galaxies with stellar masses greater than $10^{10}$\Msun. However \HI\ scaling relations are sensitive to the optical data. Scaling relations derived with SDSS data can only be reliably applied to galaxies in SDSS due to the unique photometric filters of SDSS. UV and SDSS photometry is not available for a large fraction of galaxies. Upcoming large scale \HI\ surveys in the southern hemisphere make it necessary to also establish scaling relations between other available optical data and the \HI\ content of galaxies.

In this work we derive \HI\ scaling relations for galaxies using the \HI\ Parkes All Sky Survey (HIPASS) and a variety of optical and near-infrared luminosities and diameters. We use a multi-wavelength approach to determine scaling relations between the \HI\ content of galaxies and their diameter and luminosity, in 5 and 6 optical/IR wavebands respectively. The different bands are not uniformly affected by extinction and they probe different stellar populations of a galaxy. Moreover the large sky coverage of the catalogues from which we derive our scaling relations, makes them suitable to investigate how different environments influence the HI content of galaxies. In section 2 of this paper, we describe the  \HI\ and optical datasets. In section 3, we present our scaling relation and the influence of environment on the \HI\ content. In section 4, we present applications of our scaling relations. We show that it is possible to identify individual galaxies or groups/clusters that have an HI content that deviates substantially from the expected values. Investigating \HI\ scaling relations in the southern hemisphere is especially important now, in preparation for the upcoming large \HI\ and optical surveys, such as the ASKAP \HI\ All Sky Survey, known as WALLABY (\citet{Wallaby}; \citealt{Koribalski2012}) and SkyMapper \citep{SkyMapper2007}. We present a method to estimate the outcome of a large blind \HI\ survey by predicting the \HI\ mass of 16709 galaxies in the 6dF Galaxy Survey \citep{Jones2009}. We compare these to the current HIPASS catalogues to investigate how effective our predictions are. In section 5, we summarise our results.

Throughout this paper we use $H_{0}=70$ \kms Mpc$^{-1}$.  

\FloatBarrier
\section{Data}

\subsection{HI data}

The \HI\ Parkes All-Sky Survey (HIPASS; \citealt{Barnes2001}) is a blind \HI\ survey conducted with the 64 m Parkes radio telescope covering two-thirds of the entire sky from declination $\delta = -90^{\circ}$ to +26$^{\circ}$ in the radial velocity range of $-1280 < cz < 12700$ \kms. HIPASS has a gridded beam size of 15\farcm5, a velocity resolution of 18 \kms, and an r.m.s. noise of $\sim$13 mJy. Details of the survey and of the data reduction are described in \citet{Barnes2001}. 

In our work we use the following catalogues: the HIPASS Bright Galaxy Catalog (HIPASS BGC; \citealt{BGC}), the southern HIPASS catalogue (HICAT; \citealt{Meyer2004}) with its optical counterpart (HOPCAT; \citealt{Doyle2005}), and the northern HIPASS catalogue (NHICAT; \citealt{Wong2006}) with its optical/infrared counterpart (NOIRCAT; \citealt{Wong2009}). 

The HIPASS BGC lists the \HI\ properties of the 1000 \HI-brightest extragalactic sources in the southern sky ($\delta < 0^{\circ}$, \HI\ peak flux $>$ 116~mJy). \citet{BGC} found 853 HIPASS sources associated with single optical galaxies (68 of these are marked as confused), 44 with galaxy pairs and 11 with compact groups. All but nine of the 853 single galaxies (with \vLG\ $<$ 300 \kms) are also listed in HICAT / HOPCAT. The positional accuracy of HIPASS sources is given by the gridded beam divided by the signal-to-noise ratio. Most HIPASS BGC sources have \HI\ spectra with signal to noise of at least nine, ie a position uncertainty of 1\farcm7 or better. 

HICAT covers the southern sky up to a declination of $\delta$ = +2$^{\circ}$ and contains 4315 extragalactic \HI\ sources (\vLG\ $>$ 300 \kms, \HI\ peak flux $\ga$ 40 mJy). To derive our main scaling relations we use the optical properties of the most reliable single galaxy identifications listed in HOPCAT. This sample consists of 1798 galaxies, about half of all the optical identifications by \citet{Doyle2005} and nearly twice the number of HIPASS BGC single galaxy identifications. For our study it is important to avoid using \HI\ sources with multiple or uncertain optical identifications.

In addition to our southern galaxy samples we also select a sample of northern HIPASS sources. NHICAT covers the sky from $\delta$ = +2$^{\circ}$ to 25.5$^{\circ}$ and contains 1002 extragalactic sources. Of these, 414 galaxies have reliable, single optical counterparts in NOIRCAT. This sample includes the Virgo cluster, which enables us to investigate the effect of a large galaxy cluster like Virgo on the \HI\ scaling relations. The NOIRCAT sample has corresponding SDSS data, which makes it possible to compare the difference of using SDSS \textit{r}-band photometry to non SDSS \textit{R}-band photometry.  

\subsection{Optical and infrared data} 
\label{Data}

We obtain homogeneous sets of optical and infrared properties (magnitudes and diameters) for the three HIPASS galaxy samples. These consist of magnitudes and diameters in three optical bands (\textit{B}, \textit{R}, \textit{I}) and three infrared bands (\textit{J}, \textit{H}, \textit{K}). Each individual property is drawn homogeneously from one source. In the following we briefly describe the data used for our analysis. 

For the southern galaxy samples we use the optical \textit{B}, \textit{R} and \textit{I}-band magnitudes catalogued in HOPCAT. \citet{Doyle2005} obtained these measurements from SuperCosmos plates. The catalogued \textit{B}-band magnitudes are consistent with \textit{B}-band magnitudes in other catalogues, but we find the \textit{R} and \textit{I}-band magnitudes have a systematic offset. The reason for this is that \citet{Doyle2005} measured the \textit{R} and \textit{I}-band magnitudes inside the same elliptical aperture that was used to measure the \textit{B}-band magnitudes. We investigate this systematic offset of their magnitudes and conclude that the \textit{R} and \textit{I}-band magnitudes are consistent with themselves, but need a linear scaling to be comparable with other catalogues. We scale the HOPCAT magnitudes with the following equations:
\begin{equation}
 \rmn{M}_{R \rmn{scaled}} = -2.99 +1.18~\rmn{M}_{R}, 
\end{equation}
\begin{equation}
 \rmn{M}_{I \rmn{scaled}} = -1.94 +1.06~\rmn{M}_{I}.
\end{equation}
The scaled magnitudes are in good agreement with SuperCosmos magnitudes in the 6dF Galaxy Catalogue \citep{Jones2009}. Optical \textit{B} and \textit{R}-band luminosities and diameters are good tracers of the young stellar populations in galaxies, but these bands are also very sensitive to extinction which can contribute significant measurement errors for individual galaxies. 

For a large fraction of the southern sample (1250 galaxies) we also obtain \textit{J}, \textit{H} and \textit{K}-band magnitudes from the 2MASS Extended Source Catalog \citep{Skrutskie2006}. 2MASS is currently the largest near-infrared catalog available for galaxies that covers the whole sky. The advantages of using near-infrared wavelengths are that they are less sensitive to dust obscuration and they can be used as stellar mass indicators. Furthermore, the whole sky coverage of 2MASS also makes it possible to compare our results with galaxies in the northern hemisphere. 

For the northern galaxy sample we obtain the 2MASS \textit{J}, \textit{H} and \textit{K} magnitudes directly from NOIRCAT (\citealt{Wong2009}). All other optical and infrared properties are obtained from HyperLEDA\footnote{http://leda.univ-lyon1.fr/} using only the catalogues specified in Table~\ref{tab:data}. This gives us homogeneous data for each individual band.

We correct all optical and infrared magnitudes for Galactic extinction based on \citet{Schlegel1998} and for internal absorption following \citet{Driver2008}. Galaxy diameters are also corrected for extinction following \citep{Graham2008} assuming they are entirely disc dominated.

\begin{table}
\caption{Overview of the optical and infrared data and the number of galaxies in the samples.}
\label{tab:data}
\begin{tabular}{l c c c c c c c}
\hline
 & B & R & r & I & J & H & K \\
\hline
$\delta < +2^{\circ}$ \\

magnitude & 1796 & 1796  &   -  & 1795 & 1249 & 1249 & 1250 \\
(Ref)  & (1)  &  (1)  &      &  (1) &  (2) & (2)  &  (2) \\
diameter  & 1179 &  262  &   -  & 1021 & 632  & -    & 1211 \\
(Ref)  & (3)  & (4,5) &      & (6)  & (6)  &      &  (2) \\
\hline
$\delta > +2^{\circ}$ \\
magnitude & 343  &   -   &  175 &  158 &  414 &  414 & 414  \\
(Ref)  & (3)  &       &  (7) &  (8) &  (2) &  (2) & (2)  \\
diameter  & 343  &  471  &  166 &  158 &  -   &  -   & 354  \\
(Ref)  &  (3) &  (6)  &  (7) &  (8) &      &      & (2)  \\
\hline
\multicolumn{8}{l}{(1) \citet{Doyle2005}, HOPCAT -- SuperCosmos magnitudes}\\
\multicolumn{8}{l}{(2) \citet{Skrutskie2006}, external 3$\sigma$ diameter}\\
\multicolumn{8}{l}{(3) \citet{Paturel2000}, 25 mag arcsec$^{-2}$ isophote} \\
\multicolumn{8}{l}{(4) \citet{Nilson1973}, external visual diameter}\\
\multicolumn{8}{l}{(5) \citet{Nilson1974}, external visual diameter}\\
\multicolumn{8}{l}{(6) \citet{Paturel2005}, 25 mag arcsec$^{-2}$ isophote}\\
\multicolumn{8}{l}{(7) \citet{SDSS5}, external 3$\sigma$ diameter}\\
\multicolumn{8}{l}{(8) \citet{Springob2007}, 23.5 mag arcsec$^{-2}$ isophote}\\
\hline
\end{tabular}
\end{table}

\section{Results}

\subsection{\HI\ scaling relations}
\label{scaling-relations}

To derive scaling relations we use sub samples of our main galaxy sample. To ensure that we are deriving scaling relations from an environmentally unbiased sample, we only use galaxies in low density environments. We calculate the environmental density around each of the galaxies in our main galaxy sample and only use galaxies with $\Sigma_{7} < 1$ Mpc$^{-2}$ (see details in section \ref{environment}). We also exclude galaxies with \HI\ fluxes lower than the HIPASS 95\% reliability limit ($S_{int} < 5$ Jy \kms, \citealt{Zwaan2004}) to ensure that we are using only the best quality data. These cuts exclude about 30 \% of our main galaxy sample, but improve the reliability of our scaling relations.

We investigate the effect of using volume limited samples, which result in significantly smaller sample sizes with only a few hundred galaxies. The linear regression fits to these samples are in good agreement with our non volume limited sample considering the significantly larger errors on the fitting because of the small sample size. Since the regression fitting for the volume limited sample is consistent with the non volume limited sample we decide against using a volume limited sample. We also investigate the effect of volume weighting on our sample. We find that using inverse volume weighting on our data - with the same method that is used when deriving HI mass functions - results in significantly underestimating the \HI\ mass when calculated from the optical data. The same can be seen in \cite{Toribio2011_2}. Since our aim is to establish scaling relations that can be used to predict the \HI\ content of late-type galaxies based on their optical properties we decide against using volume weighting to derive our scaling relations.

\subsubsection{Scaling relations based on magnitudes}

We determine scaling relations between the logarithm of the observed \HI\ mass of galaxies and their magnitudes in 6 different wavebands. We tested different regression fitting algorithms to determine the best method to derive our scaling relation. We find that the ordinary least-squares bisector (OLS bisector) regression line best recovers the one-to-one relation between the predicted \HI\ mass and the observed \HI\ mass. An ordinary least square (OLS) fit to the data results in an under prediction of the \HI\ mass for bright galaxies and over prediction of the \HI\ mass for faint galaxies. \cite{Feigelson1992} also recommends to use a symmetric regression line fit, such as the OLS bisector for investigating physical processes behind regressions. We use OLS bisector regression line fitting through the whole of this work. The scaling relations for the magnitudes are in the form
 \begin{equation}
 \rmn{logM}_{HI}=\alpha+\beta\cdot M_{x}
\end{equation}

where M$_{HI}$ (\Msun) is the \HI\ mass of a galaxy, $M_{x}$ is the absolute magnitude in the various bands and $\alpha$ and $\beta$ are the parameters of the relation (Table~\ref{tab:HOPCAT parameters}). The observed \HI\ mass (\Msun) is calculated with the following equation
\begin{equation}
 M_{HI}=2.356 \times 10^{5}D^{2}F_{HI},
\end{equation}
where, $F_{HI}$ (Jy) is the integrated  flux of the 21 cm emission line and $D$ is the distance of the source in Mpc using $D=v_{LG}/H_{0}$. $v_{LG}$ is the Local Group velocity calculated from the radial velocity $v_{LG}=v+300$ sin $l$ cos $b$. We use these Hubble flow distances trough the whole paper. For nearby galaxies Hubble flow distances might have large errors, but according to \cite{Zwaan2003} southern HIPASS galaxies should not be largely effected by local galaxy over-densities. We compare our calculated distances to distances derived from The Extragalactic Distance Database\footnote{http://edd.ifa.hawaii.edu/} (EDD). We find a reasonably good agreement between our calculated distances and the distances in the database. 

\begin{figure*}
\centering
\includegraphics[width=170mm]{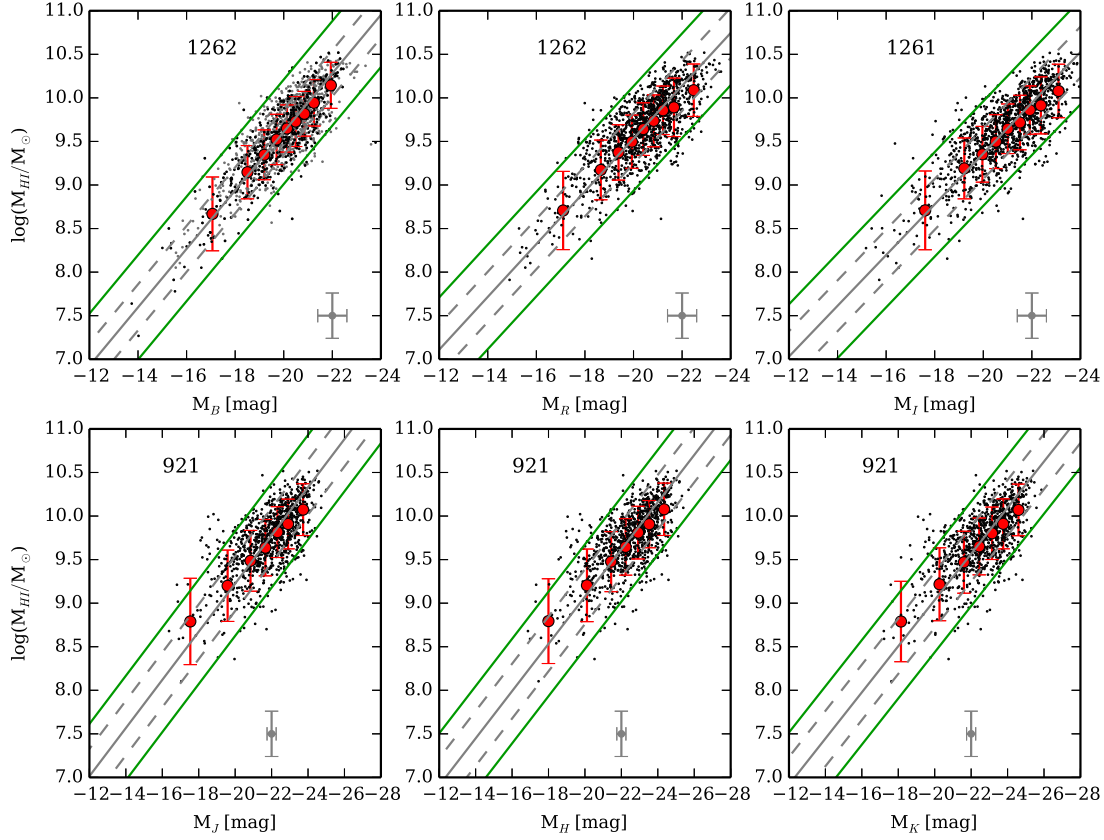}
\caption{The logarithm of \HI\ mass plotted against 6 different optical and infrared magnitudes. The solid grey line is the fitted bisector line to the data and the dashed grey lines are the 1$\sigma$ standard deviation. The red points are magnitude bins with 200 galaxies each (except for the last bin). The green lines are marking \textsc{DEF}$= \pm 0.6$, galaxies below this threshold are considered \HI-deficient and galaxies above this threshold are considered to have \HI-excess. The numbers in the top right corner of the sub-plots show the sample size in each band and the grey points with the error bars show the average uncertainty in the measurements. In the top left panel we marked the HOPCAT sample with grey points and the BGC sample with black points.}
\label{fig:mag-logHI}
\end{figure*}

Figure~\ref{fig:mag-logHI} shows our relations between the \HI\ mass of the galaxies and their magnitudes in the 6 multi-wavelength bands. The grey line shows the OLS bisector regression fit to the data. The grey dashed lines show the 1$\sigma$ standard deviation from the fitted line and the solid green lines mark deficiency factors of $\pm 0.6$ calculated from the scaling relation in each band. Galaxies outside the green lines are the most extreme outliers from our relations. Red symbols with error bars show the data binned in magnitude (each bin containing 200 galaxies, except the last bin, which consists of the remaining galaxies). We perform a bootstrap analysis for the line fitting. The fitted parameters and the bootstrap errors for all 6 bands are in Table~\ref{tab:HOPCAT parameters}.

We find a tighter correlation with the optical \textit{B}, \textit{R} and \textit{I} band magnitudes ($\sigma = 0.26, 0.29, 0.29$) than with the near-infrared \textit{J}, \textit{H} and \textit{K} bands ($\sigma = 0.31, 0.31, 0.32$). This may be because the near-infrared bands trace the old stellar population, which is usually more concentrated in the inner regions of galaxies and is not so tightly linked to the \HI\ content. The 2MASS \textit{K}-band diameter of a disk galaxy is on average 1.5-2 times smaller than it's \textit{B}-band diameter \citep{Jarrett2003}.  

\begin{figure}
\centering
\includegraphics[width=84mm]{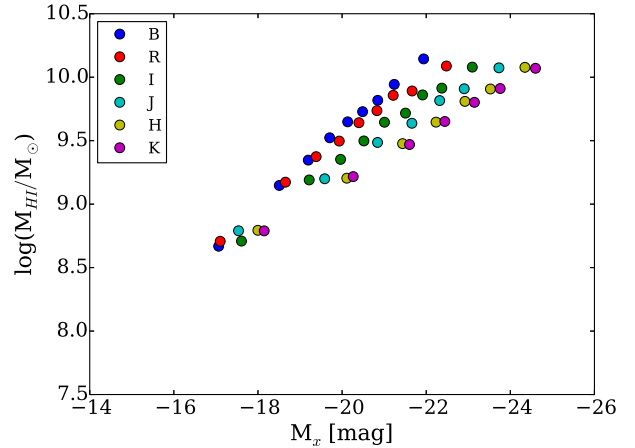}
\caption{Comparing the binned data points of the \HI\ mass - magnitude scaling relations. Different coloured points show the data in the different wavebands.}
\label{fig:compare-relations}
\end{figure}

To investigate if there is a trend between the scaling relations in the different wavebands, we compare the binned data points of the \HI\ mass - magnitudes scaling relations. We find that galaxies with a similar \HI\ mass tend to have brighter magnitudes at the longer wavelengths (Figure~\ref{fig:compare-relations}). This trend is most likely caused by the increasing dust absorption towards shorter wavelengths or the different sampled stellar populations. After binning the data in log \HI\ mass as well, we conclude that there is a small increase in the difference between the different magnitudes from the low to the high \HI\ masses, which explains the slightly different slopes of the \HI\ scaling relations, but this difference is not significant. 

For comparison we also derive scaling relations in the \textit{B}-band only using galaxies that are also in the BGC. These galaxies are the brightest galaxies in our sample with the best position accuracy and have the most certain optical identifications. We find that the scaling relation derived from this sub-sample is in good agreement with the HOPCAT sample \textit{B}-band scaling relation.

\subsubsection{Scaling relations based on diameters}
\label{scaling-relations2}

We also determine scaling relations between the observed \HI\ mass of galaxies and the logarithm of their diameter in 5 different wavebands using the same method as for the magnitudes. The scaling relations for the diameters are
\begin{equation}
 \rmn{logM}_{HI}=\alpha+\beta \cdot \rmn{log}d_{x}
\end{equation}
where M$_{HI}$ (\Msun) is the \HI\ mass of a galaxy and $d_{x}$ (kpc) is the diameter in the various bands and $\alpha$ and $\beta$ are the parameters of the relation (Table~\ref{tab:HOPCAT parameters}).

\begin{figure*}
\centering
\includegraphics[width=170mm]{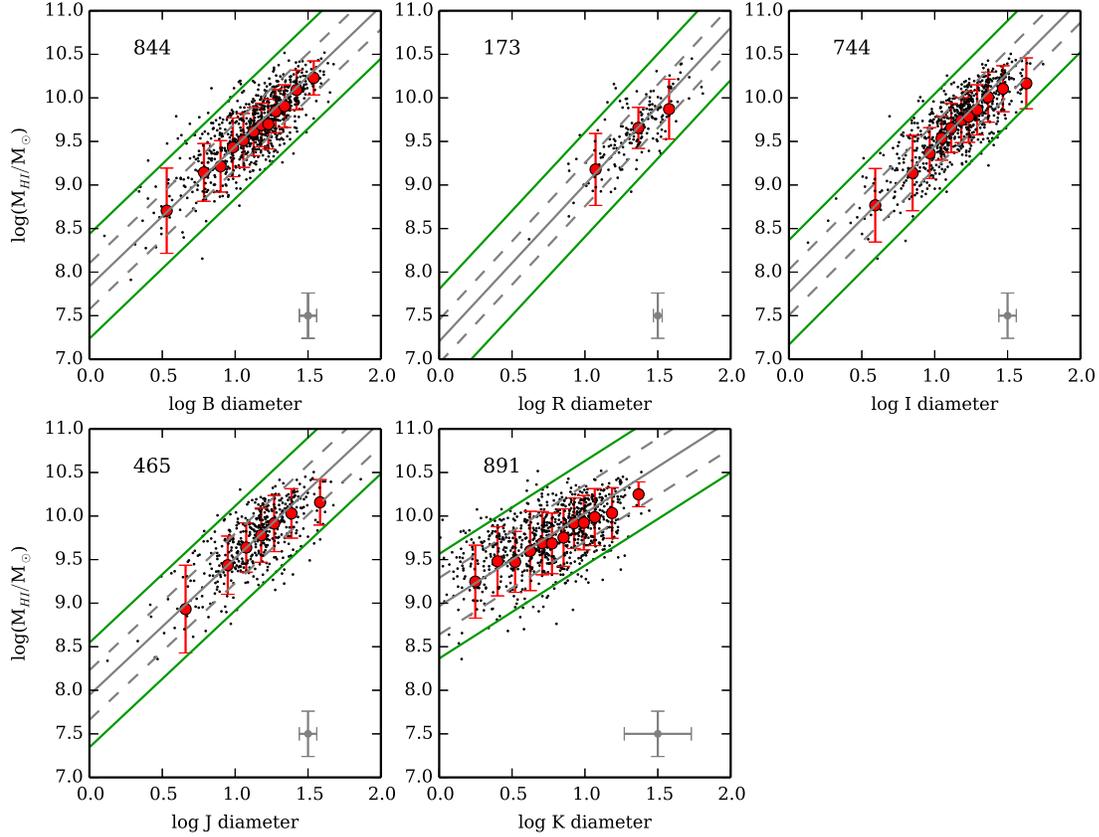}
\caption{The logarithm of \HI\ mass plotted against the logarithm of the diameter in 5 different optical and infrared bands. The solid grey line is the fitted bisector line to the data and the dashed grey lines are the 1$\sigma$ standard deviation. The red points are diameter bins with 100 galaxies each (except for the last bin). The green lines are \textsc{DEF}$= \pm 0.6$, galaxies below this threshold are considered \HI-deficient and galaxies above this threshold are considered to have \HI-excess. The numbers in the top right corner of the sub-plots show the sample size in each band and the grey points with the error bars show the average uncertainty in the measurements.}
\label{fig:logdiameter-logHI}
\end{figure*}

Figure~\ref{fig:logdiameter-logHI} shows the correlation between the \HI\ mass of the galaxies and their \textit{B}, \textit{R}, \textit{I}, \textit{J}, \textit{K} band diameters. Markings are the same as in Figure~\ref{fig:mag-logHI}. We perform a bootstrap analysis for the line fitting, the fitted parameters and the bootstrap errors for all 5 bands are in Table~\ref{tab:HOPCAT parameters}.

Similar to the magnitudes, we see tighter correlations between the \HI\ mass and the optical \textit{B}, \textit{R} and \textit{I} diameters ($\sigma = 0.25, 0.24, 0.25$) than for the near-infrared \textit{J} and \textit{K} diameters ($\sigma = 0.27, 0.29$). 

We calculate the Pearson's correlation coefficient for our relations to measure the strength of the linear dependency of the two variables (Table~\ref{tab:HOPCAT parameters}). The negative values for the magnitudes are due to the negative slope of the scaling relations for the magnitudes. This coefficient shows a similar trend to the OLS bisector fitting. The optical magnitudes and diameters show a stronger correlation than the near-infrared magnitudes and diameters. We find that our \textit{R}-band diameter scaling relation has the smallest scatter of our derived relations, however this is not necessarily due to the strength of the scaling relation. It could also be caused by the significantly smaller sample size of this band compared to the other bands. Considering the sample sizes, the scatter of the relations and the Pearson's correlation coefficient we find the \textit{B}-band magnitude \HI\ scaling relation the best to estimate the \HI\ mass of a galaxy and use it for estimating \HI\ masses hereafter.

We did not make any morphological selections for our scaling relations, but it is important to note that our scaling relations are only valid for late type galaxies, since our galaxy sample contains a relatively insignificant amount of early type galaxies (ellipticals and S0s). Our sample also contains a small number of dwarf galaxies, which do not influence the relations significantly and follow the \HI\ scaling relations derived for all the sample galaxies. 

\begin{table*}
\caption{Parameters for the \HI\ scaling relations ($\delta < +2^{\circ}$), $\alpha$ and $\beta$ are the relation parameters, N is the number of galaxies used to determine the relations and $\sigma$ is the standard deviation from the least square fit.}
\label{tab:HOPCAT parameters}
\begin{tabular}{l c c c c c c}
\hline
Type & $\beta$ & $\alpha$ & N & $\sigma$ &  &  Pearson's cofficient \\ \hline
B magnitude & -0.34 $\pm$ 0.01 & 2.89 $\pm$ 0.11  & 1262 & 0.26 & & -0.81\\
R magnitude & -0.3  $\pm$ 0.01 & 3.47 $\pm$ 0.11 & 1262 & 0.29 &  & -0.77\\
I magnitude & -0.29 $\pm$ 0.01 & 3.54 $\pm$ 0.12  & 1261 & 0.29  & & -0.77\\
J magnitude & -0.28  $\pm$ 0.01 & 3.7 $\pm$ 0.16 & 921 & 0.31 & & -0.67\\
H magnitude & -0.27  $\pm$ 0.01 & 3.65 $\pm$ 0.16 & 921 & 0.31 & & -0.67\\
K magnitude & -0.27  $\pm$ 0.01 & 3.74 $\pm$ 0.16 & 921 & 0.32 &  & -0.66\\
\hline
B-band diameters & 1.27 $\pm$ 0.04 & 8.21 $\pm$ 0.04 & 844 & 0.25 & & 0.79\\
R-band diameters & 1.39 $\pm$ 0.11  & 7.76 $\pm$ 0.15 & 173  & 0.24 & & 0.78\\
I-band diameters & 1.32 $\pm$ 0.04  & 8.17  $\pm$ 0.05 & 744 & 0.25 &  & 0.78\\
J-band diameters & 1.16 $\pm$ 0.05 & 8.41  $\pm$ 0.06 & 465 & 0.27 & & 0.77\\
K-band diameters & 0.66 $\pm$ 0.03 & 9.26 $\pm$ 0.03 & 891 & 0.29 & & 0.6\\
\hline
\end{tabular}
\end{table*}

\subsubsection{Scatter}

An important property of the scaling relations is the scatter. We investigate several effects that can broaden the scatter in the relations.

The internal extinction of galaxies may comprise one component of the scatter. We correct the magnitudes for internal extinction following \citet{Driver2008} and correct the diameters for extinction effects following \citep{Graham2008} assuming disc dominated galaxies. The scatter in the scaling relations for the magnitudes improves after the correction but we do not see a significant improvement for the diameters. This is most probably due to the large uncertainties in the corrections for internal dust and inclination. We do not find any significant trend between the \HI\ deficiency parameter and the inclination of our sample galaxies. We also find that excluding galaxies with extreme inclination ($i < 30^{\circ}$ and $i > 80^{\circ}$) does not decrease the scatter for our scaling relations.

Part of the scatter arises from environmental differences. Galaxies in high density environments have a slightly different \HI\ content compared to galaxies in low density environments. We discuss this in more detail in section \ref{environment}. After excluding galaxies in high density environments the scatter of our scaling relations decreased.

Additional factors adding to the scatter can be morphology (e.g. \citet{Haynes1984}), uncertainty of the \HI\ mass, different metallicities, different star formation histories and of course measurement errors. This means that \HI\ scaling relations always have an uncertainty that needs to be considered for their applications, such as predicting the \HI\ mass of galaxies. 

\subsection{Scaling relations from NOIRCAT}

We also derive scaling relations using galaxies in NOIRCAT \citep{Wong2009} for the \HI\ sample to investigate the effects of a large galaxy cluster, like the Virgo cluster, on \HI\ scaling relations and the difference between using SDSS or non SDSS data. Our northern sample consists of the 414 galaxies with high confidence optical counterparts in NOIRCAT. The optical and near-infrared data for this sample is derived from HyperLEDA selecting homogeneous data from single catalogues (Table \ref{tab:data}). 

We derive the correlations the same way as presented in section~\ref{scaling-relations} and present the parameters in Table~\ref{tab:NOIRCAT parameters}. The scaling relations derived from NOIRCAT are similar to the ones from the HOPCAT sample with some variations. We find that considering the sample size, the scatter of the relations and the Pearson's correlation coefficient the \textit{R}-band diameter relation is the best to estimate the \HI\ mass of a galaxy. We compare our derived scaling relations from the SDSS \textit{r}-band diameter and the \textit{R}-band diameter from \cite{Nilson1973}. We find that the average difference of the predicted log$M_{HI}$ from the \textit{r} and \textit{R}-band magnitudes is relatively small (0.02 dex), but the standard deviation is considerable (0.41 dex) and the difference for individual galaxies can be up to 2.66 dex. This shows that scaling relations derived from different optical data sets can only be used with caution to predict the \HI\ mass of a galaxy. 

\begin{figure}
\centering
\includegraphics[width=84mm]{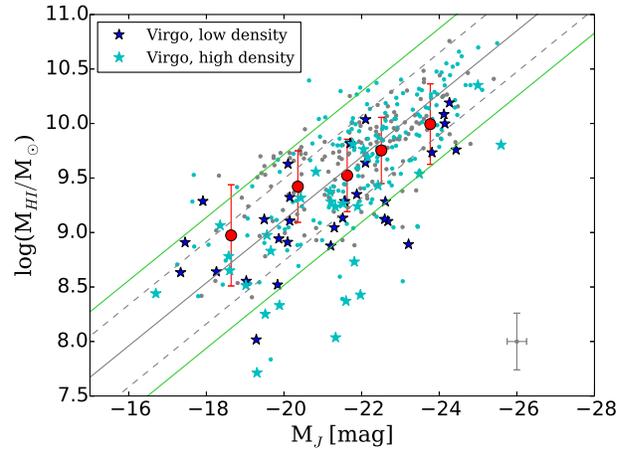}
\caption{2MASS J-band magnitude plotted against log \HI\ mass (NOIRCAT sample). Grey and cyan points mark galaxies in low and high density environments respectively. The solid grey line is the fitted bisector line and the dashed grey lines are the 1$\sigma$ standard deviation. The red points show the data in magnitude bins (40 galaxies each bin, except the last bin). The green lines are \textsc{DEF}$= \pm 0.6$. The blue and cyan stars mark galaxies in the area of the Virgo cluster.}
\label{fig:Jmag-logHI}
\end{figure}

Figure~\ref{fig:Jmag-logHI} shows our scaling relations between the \HI\ mass of galaxies in NOIRCAT and their 2MASS \textit{J}-band magnitude. We mark galaxies in cyan that are in high density environments ($\Sigma > 1$ Mpc$^{-2}$), these galaxies are not used when calculating the scaling relations. We show the \textit{J}-band magnitude, because we would like to emphasize that it is possible to identify \HI-deficient and \HI-excess galaxies not just with optical but also with near-infrared scaling relations. The northern extension of HIPASS covers the sky up to a declination of +25$^{\circ}$ including the Virgo galaxy cluster. The Virgo cluster is known to have several \HI-deficient spiral galaxies (e.g \cite{Giovanelli1983}; \cite{Haynes1986}; \citet{Chung2009}). In figure~\ref{fig:Jmag-logHI} blue and cyan stars mark galaxies in the area of the Virgo cluster (+2$^{\circ} < \delta < +20^{\circ}$, 12 h $< \alpha <$ 13 h 20 m, $v <$ 2680 \kms). There are 63 galaxies in this region with available \textit{J}-band magnitude of which 13 are among the most \HI-deficient galaxies in this sample, with at least 4 times less \HI\ than the average galaxies ($DEF > 0.6$). 

\begin{table*}
\caption{Parameters for \HI\ scaling relations ($\delta > +2^{\circ}$)T, $a$ and $b$ are the relation parameters, N is the number of galaxies used to determine the relations and $\sigma$ is the standard deviation from the least square fit.}
\label{tab:NOIRCAT parameters}
\begin{tabular}{l c c c c c c}
\hline
 Type & $\beta$ & $\alpha$ & N & $\sigma$ & &  Pearson's cofficient \\ \hline
B magnitude & -0.23 $\pm$ 0.02 & 4.91 $\pm$ 0.39 & 146 & 0.38  &  & -0.72 \\
r magnitude & -0.3 $\pm$ 0.04 & 3.43 $\pm$ 0.77 & 84 & 0.31 &  & -0.78 \\
I magnitude & -0.33 $\pm$ 0.03 & 2.59 $\pm$ 0.61 & 73 & 0.35 &  & -0.69 \\
J magnitude & -0.29 $\pm$ 0.02 & 3.34 $\pm$ 0.37 & 185 & 0.38 &  & -0.7 \\
H magnitude & -0.29 $\pm$ 0.02 & 3.23 $\pm$ 0.40 & 185 & 0.39 & & -0.69 \\
K magnitude & -0.29  $\pm$ 0.02 & 3.18 $\pm$ 0.39 & 185 & 0.39 & & -0.62\\
\hline
B-band diameters & 1.77 $\pm$ 0.07 & 7.53 $\pm$ 0.09 & 146 & 0.34 &  & 0.79 \\
R-band diameters & 1.76 $\pm$ 0.07 & 7.29 $\pm$ 0.09 & 213 & 0.29 &  & 0.81 \\
r-band diameters & 1.31 $\pm$ 0.19 & 7.81 $\pm$ 0.26 & 69 & 0.47 & & 0.56 \\
I-band diameters & 2.09 $\pm$ 0.17 & 6.83 $\pm$ 0.23 & 73 & 0.3  &  & 0.75 \\
K-band diameters & 1.47  $\pm$ 0.08 & 8.62 $\pm$ 0.07 & 166 & 0.37  &  & 0.67 \\
\hline
\end{tabular}
\end{table*}

\subsection{Environmental dependency}
\label{environment}

\HI\ scaling relations can be used to investigate the environmental effects on galaxies. We know that the \HI\ content of a galaxy is strongly influenced by the environment. For example, galaxies in high density environments tend to be on average \HI-deficient compared to galaxies in low density environments (e.g. \cite{Giovanelli1985, Solanes2001, Cortese2011}). This means that galaxies in different density environments should have slightly different scaling relations. We now investigate how our \HI\ scaling relations depend on environment. 

To quantify the environment our galaxies lie in, we calculate the galaxy surface density around each of our sample galaxies, using a reference sample from HyperLEDA. The reference sample includes all galaxies in HyperLEDA below a declination of 10$^{\circ}$ that have an apparent \textit{B} magnitude brighter than 14 mag and have measured radial velocities smaller than 6000 \kms. HyperLEDA is complete to a limiting apparent magnitude of m$_{B}$=14 mag within $\mid$b$\mid < 20^{\circ}$ and $cz \leq 6000$ \kms \citep{Giuricin2000}. We show the velocity distribution of our reference sample and the HOPCAT sample in Figure~\ref{fig:velocity-hist}. To avoid edge effects we only calculate the environment for galaxies in the HOPCAT sample with $\mid$b$\mid > 30^{\circ}$. This limit is sufficient because the average distance to the 7th nearest neighbour in our sample is $\sim 8^{\circ}$.  

\begin{figure}
\centering
\includegraphics[width=84mm]{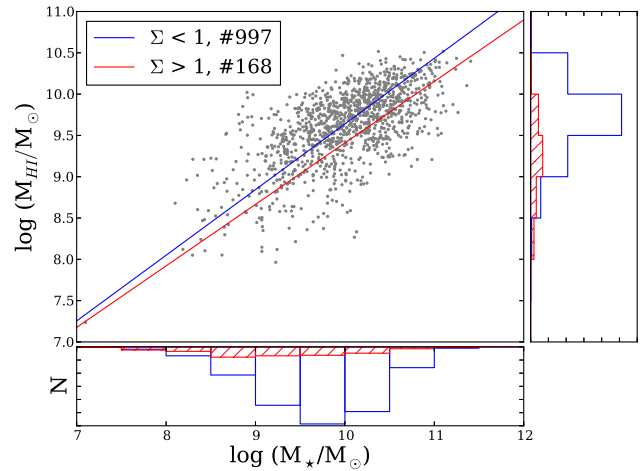}
\caption{Environment dependency of \HI\ mass as a function of stellar mass. The blue line shows the OLS bisector regression line for galaxies in low density environments ($\Sigma < 1$ Mpc$^{-2}$) and red line shows the regression line for galaxies in high density environments ($\Sigma > 1$ Mpc$^{-2}$). We show the number of galaxies in the different environments in the legend and the distribution of the sample in the histograms on the side.}
\label{fig:environment}
\end{figure}

\begin{figure*}
\centering
\includegraphics[width=170mm]{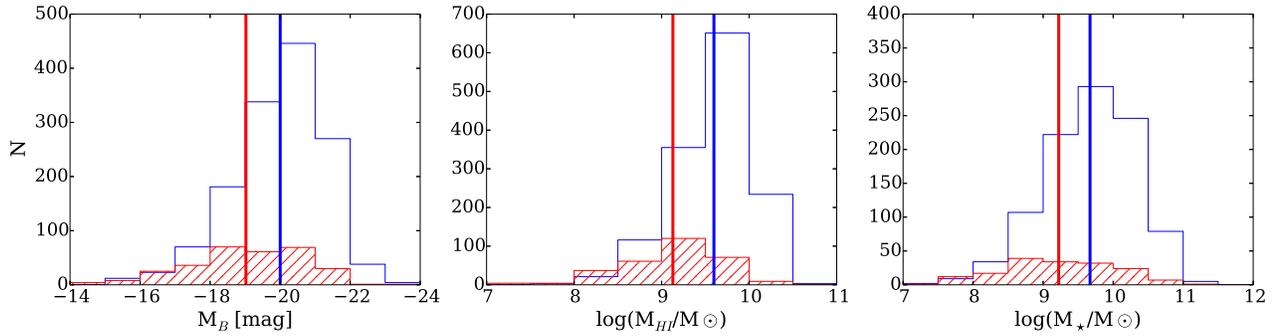}
\caption{Environment dependency of the \textit{B}-band magnitude, \HI\ mass and stellar mass of our sample galaxies. The distribution for galaxies in the low density environments ($\Sigma < 1$ Mpc$^{-2}$) is in blue and for galaxies in higher densities ($\Sigma > 1$ Mpc$^{-2}$) in red. The vertical lines show the means of the distributions.}
\label{fig:environment2}
\end{figure*}

We calculate the galaxy surface density with the projected 7th nearest neighbour method. We use a $\pm 500$ \kms cylinder centred around each galaxy and the following equation to calculate the surface density:
\begin{equation}
\Sigma _{7}(Mpc^{-2}) = \frac{7}{\pi r^{2}} 
\end{equation}
where $r$ (Mpc) is the distance to the 7th nearest neighbour. We divided the galaxies in our sample into 2 density groups: galaxies in a dense environment ($\Sigma_{7} > 1$ Mpc$^{-2}$) and galaxies in a less dense environment ($\Sigma_{7} < 1$ Mpc$^{-2}$), corresponding to an average projected galaxy density of 1 galaxy/Mpc. Since environmental effects depend strongly on stellar mass \citep{Kauffmann2004}, we show the \HI\ content in the two different environments as a function of stellar mass (Figure~\ref{fig:environment}). We find that galaxies in the higher density environment have a 0.17 dex lower average \HI\ mass than galaxies with the same stellar mass in the less dense environment. This is in good agreement with previous studies such as \citet{Cortese2011}. In figure~\ref{fig:environment2} we show the distribution of \textit{B} band magnitude, \HI\ mass and stellar mass in the two different environments. We calculate stellar masses following \citet{Bell2001} using \textit{J}-band stellar mass to light ratio and \textit{B-R} colors. We find that the peak of the distributions in the high density sample is shifted towards lower values compared to the low density sample. This suggests that late type galaxies in a low density environment are on average brighter and have a higher \HI\ mass than galaxies in high density environments.    

\section{Discusion}

\subsection{\HI\ deficiency \& excess}
\label{HI deficiency and excess}

\HI\ scaling relations are a useful tool to identify galaxies with `anomalous' \HI\ content - i.e. \HI-deficient galaxies or galaxies with \HI-excess. It can help us to understand where these galaxies are located and what is causing them to have \HI-deficiency or \HI-excess. These galaxies may have recently undergone some kind of external evolutionary processes. Their optical properties are still unchanged, but their \HI\ content is quiet different. For example we know that \HI-deficient spiral galaxies observed in the Virgo cluster have undergone recent gas stripping. \HI-excess galaxies might be accreting \HI\ from their surroundings. 

Generally a galaxy is considered to have a normal \HI\ content if it has an \HI\ deficiency parameter between -0.3 and 0.3. This corresponds to an \HI\ content 2 times more or less of an average spiral galaxy. Since the scatter of our \HI\ scaling relations is $\sim$0.3 dex, we use a more conservative limit for \HI-deficient and excess galaxies. We consider a galaxy to be \HI-excess if \textsc{DEF} $< -0.6$ and to be \HI-deficient if \textsc{DEF} $> 0.6$, which corresponds to 4 times more or less \HI\ than the average. An important test of \HI\-deficiency and \HI-excess is if we derive similar deficiency parameters in more than one waveband. This way we can avoid both individual measurement errors for a galaxy in a particular waveband and some of the errors caused by strong extinction in the shorter wavebands. It is also useful to compare the deficiency parameter calculated from the magnitude with the parameter calculated from the diameter of the galaxy. We find that the distribution of the \HI\ deficiency parameter in our samples is approximately Gaussian, centred around 0 with a standard deviation of 0.27. We find 1\% of the galaxies in our samples are \HI-excess and 4\% are \HI-deficient.

We investigate where these anomalous galaxies are located. \HI-deficient galaxies are usually found in galaxy clusters (e.g  \citealt{Giovanelli1985, Solanes2001}) and a few were found in lose galaxy groups (e.g. \citealt{Chamaraux2004, Kilborn2005, Sengupta2006}) and compact groups \citep{Verdes-Montenegro2001}. We find that about half of the galaxies that we identify as \HI-deficient are classified as members of galaxy groups or clusters in the NASA/IPAC Extragalactic Database\footnote{http://ned.ipac.caltech.edu/} (NED). Based on our density calculations, 42 \% of the \HI-deficient galaxies are in high density environments ($\Sigma > 1$ Mpc$^{-2}$) and 58 \% are in densities typical of galaxy groups or the field.

\begin{figure*}
\centering
\includegraphics[width=168mm]{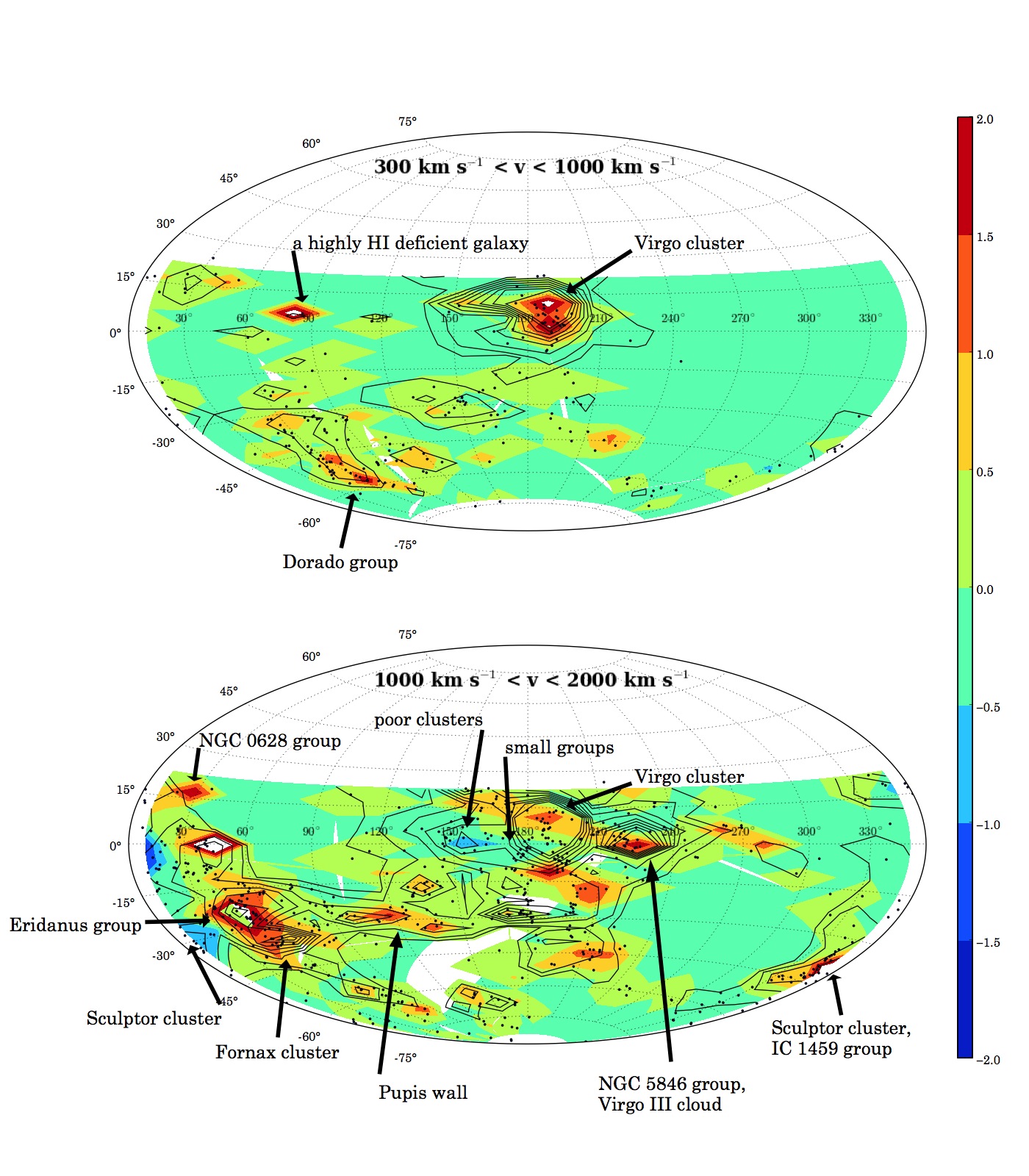}
\caption{Sky distribution of the \HI\ deficiency parameter in 20 two dimensional bins overplayed with HyperLEDA density contours. The colours represent average \HI\ deficiencies of different areas. Red and orange regions have on average more \HI-deficient galaxies and dark blue regions have on average more \HI\ rich galaxies than the green and light blue regions. Density contours are 10, 30, 50, 70, 90, 110 galaxies. Black dots represent the individual galaxies of our HOPCAT and NOIRCAT sample.}
\label{fig:sky distribution v1}
\end{figure*}

\begin{figure*}
\centering
\includegraphics[width=168mm]{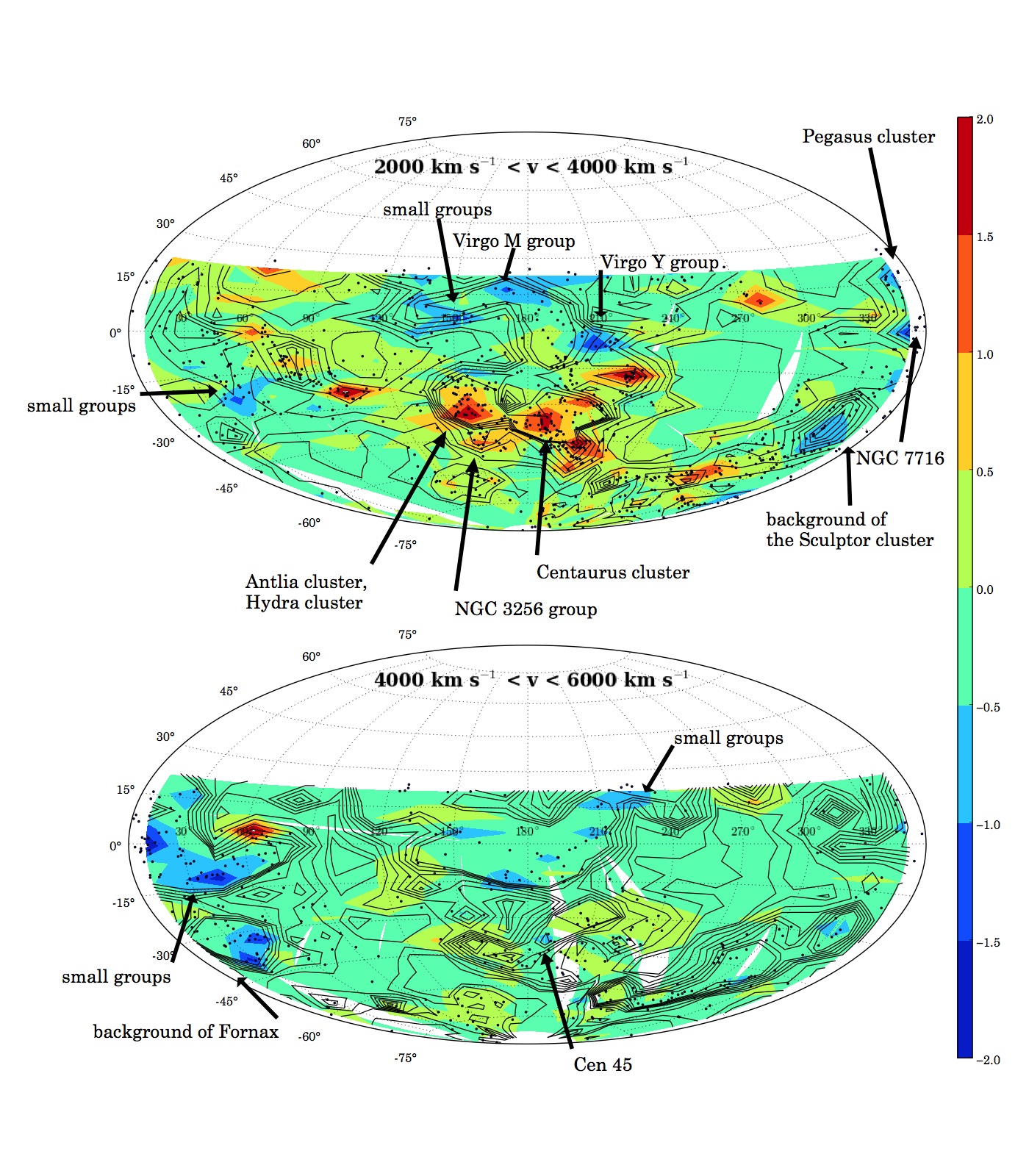}
\caption{Sky distribution of the \HI\ deficiency parameter in 20 two dimensional bins overplayed with HyperLEDA density contours. The colours represent average \HI\ deficiencies of different areas. Red and orange regions have on average more \HI-deficient galaxies and dark blue regions have on average more \HI\ rich galaxies than the green and light blue regions. Density contours are 10, 30, 50, 70, 90, 110 galaxies. Black dots represent the individual galaxies of our HOPCAT and NOIRCAT sample.}
\label{fig:sky distribution v2}
\end{figure*}

\begin{figure}
\centering
\includegraphics[width=84mm]{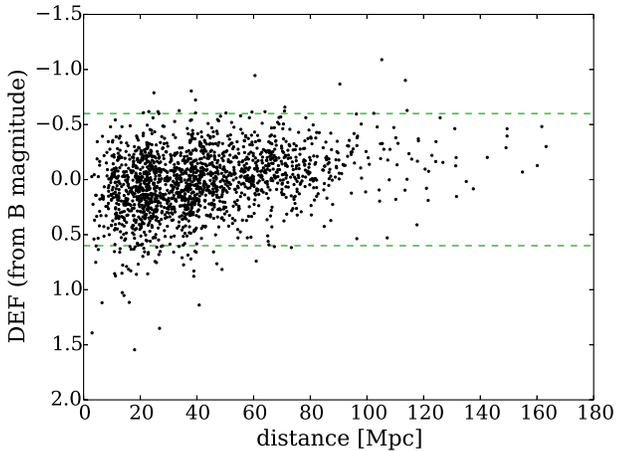}
\caption{The \HI\ deficiency factor plotted against the distance of the galaxies. The shape of this distribution is very similar to the shape of the \HI\ mass - distance plot. We can only detect \HI-deficient galaxies that are relatively close by because the survey sensitivity limit. Dashed lines mark deficiency factor of $\pm 0.6$.}
\label{fig:distance-HIdeficiency}
\end{figure}

\begin{figure*}
\centering
\includegraphics[width=168mm]{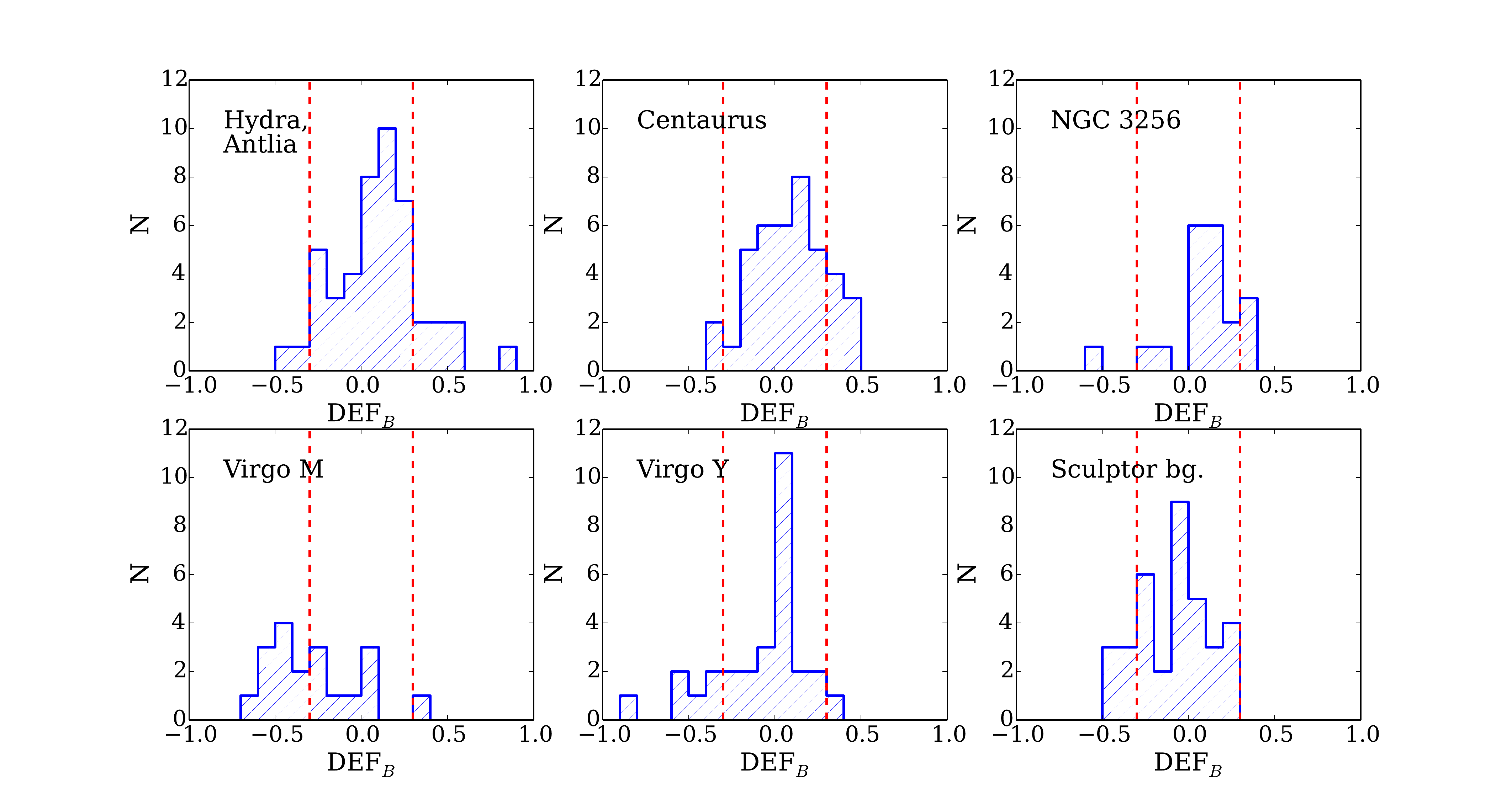}
\caption{Histograms of the \HI\ deficiency parameter for the \HI-deficient and \HI-excess regions between 2000 \kms\ $< v <$ 4000 \kms\ (first panel of Figure~\ref{fig:sky distribution v2}).}
\label{fig:HIdeficiency-histogram}
\end{figure*}

To investigate the distribution of \HI-deficient and excess galaxies on the sky we made a 2D histogram of the \HI\ deficiency parameter for the HOPCAT and NOIRCAT samples. We divide our sample into 20 spatial bins and weight each bin with the \HI\ deficiency parameter of the galaxies in it. To compare the \HI\ content with local environment density we overlay galaxy density contours from a magnitude limited HyperLEDA galaxy sample similar to the sample in section~\ref{environment}. The panels in Figures~\ref{fig:sky distribution v1} and \ref{fig:sky distribution v2} show different velocity cuts of our samples. The colour coding shows the average \HI\ deficiency, calculated from the \textit{B}-band magnitude. Orange and red colours show \HI-deficient regions and dark blue colours show \HI-excess regions. These can be interpreted as regions in the sky, that have an over density of \HI-deficient or \HI-excess galaxies compared to the average. To show this in more detail in Figure~\ref{fig:HIdeficiency-histogram} we plot the distribution of the \HI\ deficiency parameter of the galaxies in the different \HI-deficient and excess regions from the first panel of Figure~\ref{fig:sky distribution v2}. The histograms are significantly skewed. They show that there are more gas poor galaxies in the \HI-deficient regions and more gas rich galaxies in the \HI-excess regions. Figures~\ref{fig:sky distribution v1} and ~\ref{fig:sky distribution v2} show that the \HI-deficient regions tend to correlate with dens regions and \HI-excess regions seem to be on the edges of dense regions. The lower velocity segments have more \HI-deficient regions and the higher velocity segments have more \HI-excess regions - this is a selection effect that is caused by the \HI\ sensitivity limit of HIPASS. Naturally, \HI-deficient galaxies have less hydrogen than average galaxies and are harder to detect at larger distances (Figure \ref{fig:distance-HIdeficiency}). 

We can clearly see a few dominant \HI-deficient (red) and \HI-excess (blue) features in Figures~\ref{fig:sky distribution v1} and \ref{fig:sky distribution v2}. We identified the main structures. Several of these groups and clusters were previously identified as being on average \HI-deficient compared to other galaxy clusters or containing \HI-deficient galaxies (Table~\ref{tab:map-v1}), but this is the first time that regions with an abundance of gas rich galaxies were identified, such as the background of the Sculptor cluster and the outskirts of the Virgo cluster. The most prominent \HI-deficient region is the Virgo cluster with several \HI-deficient galaxies (e.g. \cite{ Giovanelli1983, Chung2009}). In Figure~\ref{fig:sky distribution v2} we can still see sub structures of the Virgo cluster with gas rich galaxies in the outskirts. Other big galaxy clusters and groups including the Eridanus group, the Fornax, Sculptor, Centaurus, Antlia and Hydra clusters. The NGC 2559 group in the Puppis wall was the first loose group of galaxies that was found to be on average \HI-deficient by \citet{Chamaraux2004}.   

Table~\ref{tab:map-v1} shows that the \HI-deficient and \HI-excess galaxies are found in a range of different density environments from large clusters to small groups and isolated galaxies. High resolution follow up observations of these galaxies can give us a valuable insight into recent galaxy evolution in different environments. These observations can help us understand which gas stripping mechanism are effective in galaxy groups and how the extremely gas rich spiral galaxies acquire their \HI. 

\begin{table*}
\caption{Previous HI deficiency and X-ray measurements for identified galaxy groups and clusters from the literature. The numbers after the group name show the population count of the group from NED.}
\label{tab:map-v1}
\begin{tabular}{l l l l}
\hline
Group name & (S+Ir):(S0+E)\% & \HI\ and group properties & X-ray measurements \\
\hline
Virgo cluster [1500]& 63:37$^{15}$  & \HI-deficient$^{1,2}$ & X-ray bright, $SUZAKU^3$,  \\
&&  & $ASCA^4$ \\
Fornax cluster (Abell 373) [58]& - & \HI-deficient$^{5,6}$& X-ray bright $ROSAT^7$  \\
&&  &\\
Dorado group (NGC 1566) [46] & 25:75$^{10}$& part of the Fornax wall$^{8,9,10}$ & $EINSTEIN^{11}$, $ROSAT^{24}$ \\
&& \HI-deficient group members$^{8}$ & \\
Pupis wall [-]& - & a chain of loose groups$^{14}$ & - \\
& & first \HI-deficient galaxies in groups$^{14}$ &\\
Pegasus cluster [6] & 82:18$^{15}$& \HI-deficient$^{2,15}$ & weak X-ray emission$^{15} ROSAT^{16}$  \\
&&  & \\
Centaurus cluster (Abell 3526) [100] & - & normal \HI\ content$^2$ & $ROSAT^{7,17}$  \\
&&   \\
Antlia cluster (Abell 636) [30] & - & - & $ROSAT^{17,18}$ \\
&&\\
Hydra cluster (Abell 1060) [107]& - & \HI-deficient$^2$ & $ROSAT^{7,17}$, $SUZAKU^{3}$ \\
&& HI rich dwarf galaxies$^{19}$ & $XMM-Newton^{20}$ \\
Eridanus group [39]& 54:46$^{15}$ & \HI-deficient galaxies$^{22}$ & X-ray measurment$^{22}$ \\
&&  & \\
IC 1459 group (Sculptor Cluster) [16]& 71:29$^{10}$ & normal group \HI\ content$^{9, 10, 23}$   & $ROSAT^{24}$ \\ 
&& \HI-deficient group members &\\
NGC 5846 group [5] & early-type& \HI-deficient$^{27}$ &  X-ray bright $ROSAT^{24}$, \\
& dominated$^{27}$ & & $CHANDRA^{25}$ \\
NGC 0628 group [7]&  late-type & normal group \HI\ content$^{23}$ & no observations$^{23}$\\
& dominated$^{23}$ & \HI-deficient group member$^{23}$& \\
NGC 7716 group [5] & late-type & normal group \HI\ content$^{23}$& X-ray faint group$^{23}$\\
& dominated$^{23}$ &    &   \\
NGC 3256 [10]& - & several peculiar and interacting galaxies$^{26}$ & -\\
&& &  \\
\hline
\multicolumn{2}{l}{$^1$\citet{Giovanelli1985}}&{$^{15}$\citet{Levy2007}}\\
\multicolumn{2}{l}{$^2$\citet{Solanes2001}}&{$^{16}$\citet{Mahdavi2000}}\\
\multicolumn{2}{l}{$^3$\citet{Shang2009}}&{$^{17}$\citet{Reiprich2002}}\\
\multicolumn{2}{l}{$^4$\citet{Shibata2001}}&{$^{18}$\citet{Ikebe2002}}\\
\multicolumn{2}{l}{$^5$\citet{Waugh2002}}&{$^{19}$\citet{Duc1999}}\\
\multicolumn{2}{l}{$^6$\citet{Schroder2001}}&{$^{20}$\citet{Piffaretti2011}}\\
\multicolumn{2}{l}{$^7$\citet{Eckert2011}}&{$^{21}$\citet{Brough2006}}\\
\multicolumn{2}{l}{$^8$\citet{Kilborn2005}}&{$^{22}$\citet{Omar2005}}\\
\multicolumn{2}{l}{$^9$\citet{Kilborn2009}}&{$^{23}$\citet{Sengupta2006}}\\
\multicolumn{2}{l}{$^{10}$\citet{Brough2006}}&{$^{24}$\citet{Osmond2004}}\\
\multicolumn{2}{l}{$^{11}$\citet{Jones1999}}& {$^{25}$\citet{Machacek2011}}\\
\multicolumn{2}{l}{$^{12}$\citet{Westmeier2011}}&{$^{26}$\citet{English2010}}\\
\multicolumn{2}{l}{$^{13}$\citet{Westmeier2013}}&{$^{27}$\citet{Haynes1991}}\\
\multicolumn{2}{l}{$^{14}$\citet{Chamaraux2004}}&\\
\hline
\end{tabular}
\end{table*}

\subsection{Predicting future \HI\ surveys}

In this section we explore how \HI\ scaling relations may be used to estimate the \HI\ content and detectability of (late-type) galaxies from an optical redshift survey. In principle, the derived relations should help in predicting the outcome of future blind \HI\ surveys, like WALLABY \citep{Wallaby}, but in practise there are numerous difficulties. While \HI\ surveys predominantly detect nearby star-forming galaxies, optical redshift surveys generally target magnitude-limited samples of bright, compact galaxies.

\begin{figure}
\centering
\includegraphics[width=84mm]{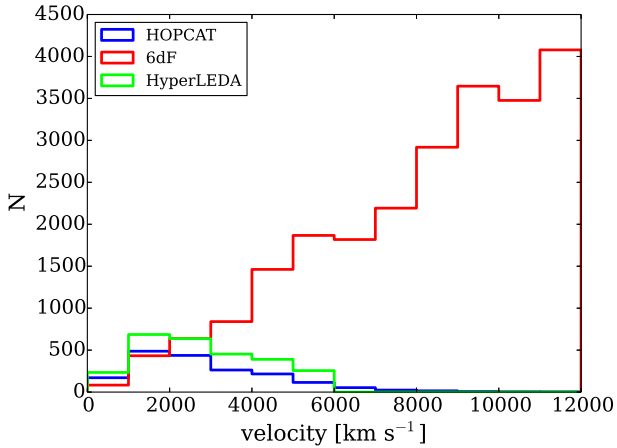}
\caption{Velocity distribution of the optically identified galaxies in the southern HIPASS sample compared to galaxies in 6dFGS and galaxies from HyperLEDA ($\delta < 10^{\circ}$ and M$_{B} < 14$ mag).}
\label{fig:velocity-hist}
\end{figure}

We use the 6dF Galaxy Survey (6dFGS, \citet{Jones2009}), a near-infrared selected redshift and peculiar velocity survey targeting galaxies in the southern sky. 6dFGS provides reliable SuperCOSMOS \textit{B} and \textit{R} magnitudes - also used to characterise the optical properties of HIPASS galaxies \citep{Doyle2005} -  for over 10 000 galaxies in the HIPASS volume ($300 < v < 13000$ \kms, $\mid$b$\mid < 20^{\circ}$). The velocity distributions of 6dFGS and HIPASS galaxies, shown in Figure~\ref{fig:velocity-hist}, differ significantly with median redshifts of 0.053 and 0.009, respectively, suggesting that only very few of the nearby 6dFGS galaxies can be detected in HIPASS. 

In order to make our \HI\ mass predictions we use the following properties from the 6dFGS catalogue: local group velocities, SuperCOSMOS \textit{B}-band magnitudes. We only select objects that are classified as galaxies and have good quality redshifts. We correct all the magnitudes obtained from the 6dF catalogue for galactic extinction and inclination effects following the same methods as described in section~\ref{Data}. Subsequently, there are 16709 galaxies in our 6dF sample, for which we calculate their predicted \HI\ mass using our \textit{B} magnitude scaling relation. 

To compare our predictions with catalogued galaxies in HIPASS, we apply a `detection limit' and a peak flux cut to the calculated \HI\ masses. We use the 95\% reliability level of the HIPASS catalogue (HICAT), which is at an integrated flux of 5 Jy \kms\ \citep{Zwaan2004}. This gives a detection limit of 1.179$\times 10^{6} D^{2}$ \Msun where $D$ (Mpc) is the distance to the galaxy. A reliable \HI\ detection also requires the peak flux of the galaxy to be 3$\sigma$ above the noise level of HIPASS. We calculate the expected peak fluxes for the 6dF galaxies by using the \textit{B}-band Tully-Fisher relation to estimate their velocity width. Using this and their predicted integrated flux we calculate the peak flux for our sample as well. After applying the integrated flux detection limit and the 3$\sigma$ peak flux cut 749 galaxies remain in our sample.

\begin{figure*}
\centering
\includegraphics[width=170mm]{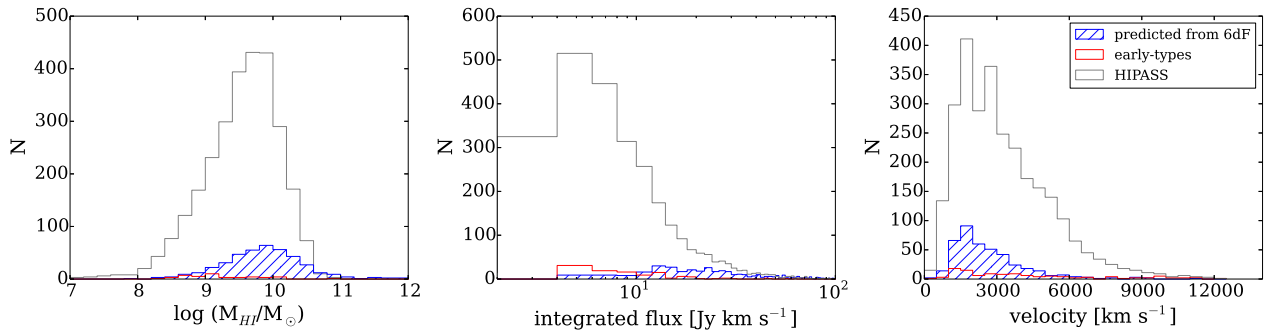}
\caption{Distribution predicted \HI\ properties of 6dF galaxies from \textit{B}-band magnitudes (blue hatched) compared to HIPASS (grey, $\mid$b$\mid > 20^{\circ}$). Red steps represent early-type galaxies (9\%) remaining in the sample after the concentration index cut.}
\label{fig:predicted HI}
\end{figure*}

Further analysis showed that a significant fraction of this sample are elliptical galaxies (30 \%), based on de Vaucouleurs morphology classifications from HyperLEDA. Whilst a number of early-type galaxies contain significant amounts of \HI\ (e.g. \citealt{Sadler2002, Oosterloo2007, Serra2012}), our scaling relations are mainly derived from late type galaxies. Thus they are not suited to estimate the \HI\ mass of early type galaxies. We investigated how to exclude early type galaxies from our sample, where morphological classifications are not available. The available 6dFGS colours (\textit{B-R}, \textit{B-J} etc.) are not suitable for a colour cut, as there is only a small offset between red and blue galaxies in the optical colour-magnitude diagrams \citep{Proctor2008}. UV-optical colours would be needed for an efficient separation of the blue and red sequence  \citep{Strateva2001}. The concentration index is an indicator of galaxy morphology and can be used to separate galaxies in the blue cloud and the red sequence \citep{Driver2006, Baldry2006}. Thus we use a 2MASS K-band concentration index cut at K$_{conc} < 3.5$ to exclude the majority of early type galaxies from our predictions. After applying this cut we predict 410 6dFGS galaxies to be detectable in HIPASS. Figure~\ref{fig:predicted HI} shows the predicted \HI\ mass, integrated \HI\ flux and the velocity distribution of these galaxies. Properties of 6dFGS galaxies are in blue and we mark the known early type galaxies in red. We still have a small fraction of early type galaxies (11 \%) in our sample. These are galaxies that passed the 2MASS concentration index cut, but for which morphological classifications from HyperLEDA indicates that they are ellipticals. We also show the same properties of HICAT galaxies in the same volume ($\mid$b$\mid < 20^{\circ}$) in grey. The predicted \HI\ mass and integrated \HI\ flux distributions have a similar shape to the distributions of HIPASS. Both \HI\ mass distribution peak around \HI\ mass ($10^{9.8}$\Msun). The lack of predicted low mass galaxies is due to the lack of low velocity galaxies in 6dF (Figure~\ref{fig:predicted HI}). 

\begin{figure}
\centering
\includegraphics[width=84mm]{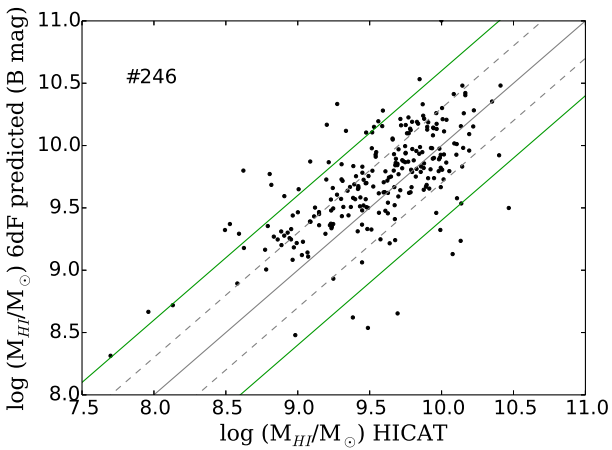}
\caption{Comparison of the predicted \HI\ mass from 6dF \textit{B}-band magnitude against the observed \HI\ mass from HICAT. The solid grey line is the 1-1 line, the dashed lines mark $\pm$0.3 dex and the green lines show $\pm$0.6 dex.}
\label{fig:compare-predicted-observed1}
\end{figure}

We compare the galaxies predicted to be detectable in HIPASS to the HIPASS source catalogue HICAT and find 246 (60 \%) galaxies matching. (We match coordinates within 15' and $\pm 300$ \kms). We compare the predicted to the measured \HI\ masses of the matched galaxies and find a good agreement, considering the scatter of our scaling relations (Figure~\ref{fig:compare-predicted-observed1}). The standard deviation between the logarithm of the observed and the predicted \HI\ mass is 0.44 dex. This is slightly larger than the standard deviation for our scaling relations, but considering the difference in sample size it is in good agreement with our scaling relations.

There are 164 galaxies that we predict to be detectable in HIPASS, but are not in HICAT. We examine the HIPASS data cubes at the positions of these galaxies and measure their \HI\ fluxes using the {\sc Miriad} \citep{Miriad} routine {\sc mbspec}. We estimate the profile width for the flux integration using the Tully-Fisher relation for the \textit{B}-band absolute magnitude. We are able to measure the integrated flux for 101 galaxies (61\%) and find an average \HI\ mass of $1.33 \times 10^{10}$ \Msun. After visual examination of the derived spectra we conclude that some of these galaxies show a typical \HI\ profile in the data cubes, but they did not qualify for the source catalogue because of low signal to noise, strong baseline ripples or RFI. A detailed investigation of these galaxies may yield additional HIPASS detections, but is beyond the scope of this work. We are unable to measure an \HI\ flux for 63 galaxies. Possible reasons for this can be poor quality \HI\ data; poor quality photometric data used for the predictions; over prediction of the \HI\ mass for a class of galaxies; early-type galaxies in the sample or \HI-deficient galaxies.

We conclude that it is possible to get a good estimate of the outcome of future \HI\ surveys using \HI\ scaling relations and an optical redshift survey. Using data from the 6dFGS redshift survey and 2MASS we predicted 410 galaxies to be detectable in HIPASS, which is about 10\% of the galaxies in the HIPASS catalogue. This is a reasonable outcome considering the significantly smaller number of nearby galaxies in 6dFGS and the different galaxy populations of the two surveys. 6dFGS is more sensitive to early-type gas poor galaxies whereas HIPASS is sensitive to late-type gas rich galaxies. To predict a whole blind \HI\ survey, reliable optical photometry, redshift and basic morphology data is needed for all galaxies in the volume. \HI\ scaling relations can also aid the fast identification of unusual sources and the investigation of detection limits.

\section{Summary}

We derive new multi-wavelength \HI\ scaling relations for galaxies using the \HI\ Parkes All Sky Survey (HIPASS) and a variety of optical and near-infrared luminosities and diameters. We find the \textit{B}-band scaling relations have the lowest scatter and therefore are the best to predict the \HI\ mass of a galaxy. The scaling relations in all wavebands are sufficient to give a good estimation of \HI\ mass for late-type galaxy. 

We investigate the environmental dependency of the \HI\ content of galaxies in two different environment densities. We find that galaxies in the high density environment tend to have on average less \HI\ than galaxies with the same stellar mass in the low density environment. We also find that galaxies in the high density environment tend to have a smaller stellar mass and lower luminosity than galaxies in the low density environment. 

\HI\ scaling relations are useful tools to identify galaxies with anomalous \HI\ content, i.e. \HI-deficient and \HI-excess galaxies. These galaxies may have been affected by recent gas stripping or gas accretion. We find 4\% of the galaxies in our samples to be \HI-deficient and 1\% to have \HI-excess. We find that about half of the galaxies that we identify as \HI-deficient are classified as members of galaxy groups or clusters in NED.

We map the global distribution of \HI-deficient and \HI-excess galaxies on the sky and compare it to the large scale structure of galaxies. We find several regions on the sky that have an over density of \HI-deficient and \HI-excess galaxies compared to the average. The \HI-deficient regions correlate with dense regions on the sky and the \HI-excess regions align with the edges of galaxy clusters and groups. This is the first time regions with an abundance of \HI\ rich galaxies were identified. We also identify the main galaxy groups and clusters that are aligned with the \HI-deficient and \HI-excess regions.

We show the potential of using \HI\ scaling relations to predict future \HI\ surveys based on an optical redshift survey. We predict the \HI\ mas of 16709 galaxies from the 6dF redshift survey and and find that 410 galaxies should be detectable in HIPASS. 60\% of these galaxies are in the HIPASS catalogue. We conclude that it is possible to get an estimate of the outcome of future \HI\ surveys using \HI\ scaling relations and an optical redshift survey. To predict a whole blind \HI\ survey, reliable optical photometry, redshift and basic morphology data is needed for all galaxies in the volume. \HI\ scaling relations can aid the fast identification of \HI-deficient and \HI-excess sources and the investigation of detection limits.

Investigating \HI\ scaling relations in the southern hemisphere is especially important now, in preparation for the upcoming large \HI\ and optical surveys, such as the ASKAP \HI\ All Sky Survey, known as WALLABY (\citet{Wallaby}; \citealt{Koribalski2012}), SkyMapper \citep{SkyMapper2007} and future surveys with the Square Kilometre Array (SKA). 

\section{Acknowledgements}

We would like to thank Barbara Catinella and Ivy Wong for their very helpful comments and conversations.

We would also like to thank the anonymous reviewer for the comments and suggestions, which helped us to significantly improve this paper. 

The Parkes telescope is part of the Australia Telescope which is funded by the Commonwealth of Australia for operation as a National Facility managed by CSIRO. 

We acknowledge the usage of the HyperLeda database (http://leda.univ-lyon1.fr).

This publication makes use of data products from the Two Micron All Sky Survey, which is a joint project of the University of Massachusetts and the Infrared Processing and Analysis Center/California Institute of Technology, funded by the National Aeronautics and Space Administration and the National Science Foundation.

This research has made use of the NASA/IPAC Extragalactic Database (NED) which is operated by the Jet Propulsion Laboratory, California Institute of Technology, under contract with the National Aeronautics and Space Administration. 

Funding for the SDSS and SDSS-II has been provided by the Alfred P. Sloan Foundation, the Participating Institutions, the National Science Foundation, the U.S. Department of Energy, the National Aeronautics and Space Administration, the Japanese Monbukagakusho, the Max Planck Society, and the Higher Education Funding Council for England. The SDSS Web Site is http://www.sdss.org/.

The SDSS is managed by the Astrophysical Research Consortium for the Participating Institutions. The Participating Institutions are the American Museum of Natural History, Astrophysical Institute Potsdam, University of Basel, University of Cambridge, Case Western Reserve University, University of Chicago, Drexel University, Fermilab, the Institute for Advanced Study, the Japan Participation Group, Johns Hopkins University, the Joint Institute for Nuclear Astrophysics, the Kavli Institute for Particle Astrophysics and Cosmology, the Korean Scientist Group, the Chinese Academy of Sciences (LAMOST), Los Alamos National Laboratory, the Max-Planck-Institute for Astronomy (MPIA), the Max-Planck-Institute for Astrophysics (MPA), New Mexico State University, Ohio State University, University of Pittsburgh, University of Portsmouth, Princeton University, the United States Naval Observatory, and the University of Washington.

\bibliography{paper1}

\end{document}